\documentstyle[prl,aps,epsf,multicol]{revtex}
\begin{document}
\draft

\title{Quantum Confinement Transition in a d-wave Superconductor}
\author{C. Lannert$^1$, Matthew P. A. Fisher$^2$, and T. Senthil$^2$}
\address{$^1$Department of Physics, University of California, Santa
Barbara, CA 93106 \\ 
$^2$Institute for Theoretical Physics, University of California, 
Santa Barbara, CA 93106--4030}  

\date{\today}
\maketitle

\begin{abstract}
We study the nature of the zero-temperature phase transition between a
d-wave superconductor and a Mott insulator in two dimensions. In this
``quantum  confinement transition'', spin and charge are confined to
form the electron in the Mott insulator.
Within a dual formulation, direct transitions from d-wave
superconductors at half-filling to insulators with spin-Peierls (as
well as other) 
order emerge naturally. The possibility of \emph{striped 
superconductors} is also discussed within the dual formulation.
The transition is described by nodal fermions
and bosonic vortices, interacting via a long-ranged statistical
interaction modeled by two, coupled Chern-Simons gauge fields, and
the critical properties of this model are discussed.
\end{abstract}
\vspace{0.15cm}


\begin{multicols}{2}
\narrowtext 

\section{introduction}

\begin{figure}
\epsfxsize=3.5in
\centerline{\epsffile{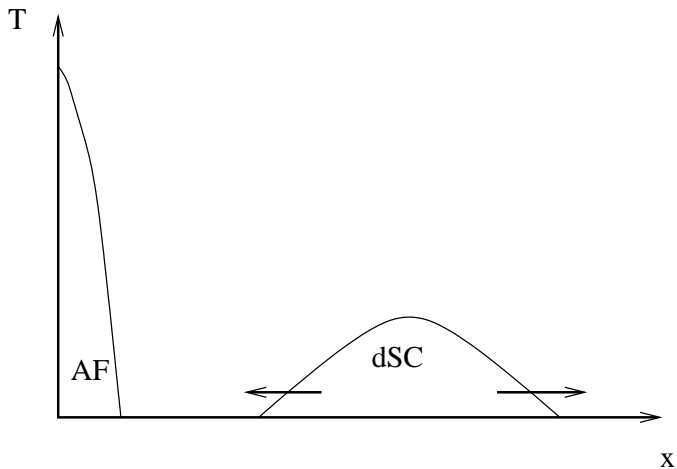}}
\vspace{0.15in}
\caption{Schematic phase diagram for the high $T_c$ cuprates.}
\vspace{0.15in}
\label{phasediagram}
\end{figure}

In recent years, due to remarkable experimental progress \cite{rev},
the cuprate superconductors have revealed a host of
mysterious phases as their chemical doping is varied. Indeed, it would
seem as though these materials exhibit many of the wide range of behaviors
possible for low-dimensional, highly-correlated electron
systems. Centrally located within the phase diagram and adjacent to
many of these puzzling regions is the d-wave superconductor. Beginning
in this well-understood phase, one may develop theoretical descriptions of
other, non-superconducting phases. Of particular interest are the
$T=0$ quantum phases, both in the very underdoped and heavily
overdoped regimes. The schematic situation is shown in
Fig.\ref{phasediagram}. 

When describing a 2-dimensional superconductor, topological
defects in the Cooper pair 
wavefunction (BCS vortices) are of particular importance. Being
bosonic, once they proliferate, they condense at $T=0$,
destroying superconductivity \cite{WZ} . 
In this way, a description of quantum phases with strong
pairing correlations but lacking the phase coherence that is
superconductivity emerge quite naturally as \emph{vortex condensates}
\cite{BFN}.
If the superconductor is d-wave, there is the additional complication
of low-energy quasiparticles.
As recently emphasized \cite{Z2}, there is a statistical
interaction between these spin-carrying quasiparticles and the
vortices, making the resulting theory strongly-interacting.

Singlet-paired superconductors can be recast in a spin-charge
separated form \cite{KR}: the condensate carries all the charge but
no spin, while the quasiparticles are electrically neutral 
with spin $1/2$. Most other
well-understood phases of electrons (such as the Fermi liquid) are
spin-charge confined.
It was recently argued \cite{QCT} that many puzzling aspects of the
cuprate phase diagram could be understood in terms of the
fractionalization and confinement of electrons.
In this approach, the regions containing the pseudo-gap
(and superconducting) phase are characterized by the presence of
spin-charge separation (electron fractionalization) \cite{confin},
while the heavily overdoped regions are
spin-charge confined. Between the two, a quantum confinement
transition might cause critical behavior
over wide regions of the high $T_c$ phase
diagram. A key feature of the cuprates is the close proximity between
a d-wave superconducting phase and a Mott insulating phase. Here, we
work at half-filling and look at direct transitions between these two
phases, viewing them as confinement transitions.
We seek to answer two broad questions regarding the nature of such a
transition. 
 
First, in terms of phenomenology, what sorts of states might
we find when separate spin- and charge- carrying excitations are
confined to form electrons? As we shall see, a remarkable feature of
superconductivity with one electron per unit cell is that in the dual
theory, the vortices are fully frustrated. When the vortices
proliferate and condense, this frustration leads to the existence of
multiple vortex condensates which break spatial symmetries. In
particular, we find vortex condensates  
which destroy superconductivity (and at half-filling, describe
Mott insulators) as well as vortex condensates which preserve
superconductivity. As a consequence of the vortex
frustration, we find direct transitions from 
superconducting states to insulating states which
spontaneously break rotational and/or translational symmetries, as
well as the existence of superconducting states which have non-trivial
spatial structure.
Although we work at exactly one electron per unit cell, where the vortex
theory is \emph{fully} frustrated, it is hoped that even away from
half-filling the qualitative features of our results will remain
valid, in particular, the tendency toward spatial modulation near
half-filling.  In general, we hope that our explorations of frustrated
vortex systems can yield insights into quantum phases of electrons
which are complicated by the presence of competing interactions.

Second, as a specific example, we look at the critical properties of the
the confinement transition between a spatially-modulated d-wave
superconductor and a Mott insulator with the same broken translational
symmetry. Characterized by the presence of long-ranged statistical 
interactions, which affect the confinement of spin and charge, this
quantum critical point should
have interesting universal properties. Within a special region of
parameter space, we explore this transition analytically using
renormalization group (RG) methods.

Before we begin to address these questions, we first lay out the
basics of the model under consideration. This model was introduced in
Ref. \cite{Z2} and many of its justifications and consequences can be found
therein. Here, we provide only a whirlwind tour of its derivation
and usefulness.

\section{the model: $Z_2$ gauge theory}

We begin by formally writing the electron creation operator as a
product of \emph{two} operators, one of which carries the spin of the
electron and the other, the charge. These operators are defined with
the singlet-paired superconductor in mind. If we write the Cooper pair
creation operator as $e^{i\varphi_r}$, we construct our spinless
charge e boson (called a ``chargon'') from the Cooper pair as:
\begin{equation}
b_r^{\dag} = s_r e^{i\varphi_r /2} \equiv e^{i\phi_r}, s_r=\pm 1.
\end{equation}
The chargon is ``half a Cooper pair'' in the sense that the \emph{square}
of $b_r^{\dag}$ creates a Cooper pair. The neutral spin one-half
particle (called a ``spinon'') is obtained by removing the charge from
the electron: 
\begin{equation}
f_{r\alpha}^{\dag} = b_r c_{r\alpha}^{\dag} .
\end{equation}
As we shall see shortly, this spinon can be thought of as a
neutralized BCS quasiparticle.
With these definitions, we may perform a change
of variables in a suitable Hamiltonian describing electrons and Cooper
pairs, resulting in a theory of chargons and 
spinons. However, the Hilbert space of chargons and spinons is
much larger than that of electrons; for instance, the state with a
single spinon but no chargons can be written down, but this state is
unphysical and should be removed from the working Hilbert space. In
other words, we may make this change of variables only if we
additionally impose a \emph{constraint} that the sum of the number of
chargons, $N_r$ (canonically conjugate to the chargon phase,
$[\phi_r,N_{r'}] = i \delta_{rr'} $ ),
and the number of spinons, 
$\rho_r=f_{r\alpha}^{\dag}f_{r\alpha}$, on each site is an even
integer:
\begin{equation}
(-1)^{N_r+\rho_r} = 1.
\end{equation}
This constraint can be implemented within a Euclidean path integral
representation, resulting in a theory of spinons and chargons coupled
to a $Z_2$ gauge field \cite{Z2}. It should be noted
that the constraint used here is not the same as Gutzwiller
projection, and does not disallow doubly-occupied sites. 

For an odd
number of electrons per unit cell and d-wave pairing correlations, the
appropriate action in the $Z_2$ gauge theory is:
\begin{eqnarray}
\label{ElecTheory}
S &=& S_c + S_s + S_B, \\
S_c &=& -t_c\sum_{<ij>} \sigma_{ij} (b^*_ib_j + h.c.) , \\
S_s &=& - \sum_{<ij>} \sigma_{ij} (t^s_{ij}\overline{f}_{i}
f_{j} + t^{\Delta}_{ij} f_{i\uparrow}f_{j\downarrow} +c.c.) \nonumber \\
  & &  - \sum_i \overline{f}_{i} f_{i}  ,\\
S_B &=& -i\frac{\pi}{2} \sum_{i,j=i-\hat{\tau}} (1-\sigma_{ij}) ,
\end{eqnarray}
where i and j label sites on a cubic space-time lattice. The Ising
gauge field minimally coupled to the chargons and spinons, 
$\sigma_{ij}$, can take values $\pm 1$, and $S_B$ is a Berry's phase
term. 

One may arrive at this action by making the above-mentioned change of
variables in a Hubbard-type Hamiltonian, as described in
Ref. \cite{Z2}. Alternatively, this model can be taken as a starting
point for describing systems with local singlet pairing correlations
as well as Mott insulating tendencies.
To exhibit the reasonableness of this model, consider the limits of
infinite and vanishing $t_c$. For $t_c \rightarrow \infty$, the
bosonic chargons will condense and the $Z_2$ gauge field will become
frozen with $\sigma_{ij} =1$, which frees the spinons. This phase is
simply the d-wave superconductor. The action reduces to S = $S_s$,
which is just the Bogoliubov-deGennes action, with the spinons
becoming the BCS d-wave quasiparticles.  
In the opposite limit, $t_c
\rightarrow 0$, the chargons are gapped into an insulating state. At
$t_c = 0$, the chargons may be trivially integrated out. The
remaining action is just $S = S_s +S_B$. It is shown in
Ref.\cite{Z2} that the partition function for this remaining spin
theory is formally equivalent to that of the Heisenberg
antiferromagnetic spin 
model. Therefore, we see the attractiveness of this model for the
cuprate system, which also exhibits both superconductivity and
antiferromagnetism.
Many other additional properties of this action between these two
limits are elucidated in Ref. \cite{Z2}, in
particular, the presence of both spin-charge \emph{confined} and
\emph{deconfined} phases \cite{expU1}. 

The charge sector in Eqn. \ref{ElecTheory} is described in terms of
the bosonic chargons, minimally coupled to a $Z_2$ gauge field. In two
spatial dimensions, vortices in the boson many-body wavefunction are
point-like. This allows for a particularly elegant dual description
where the vortex rather than the chargon is the fundamental degree of
freedom. In this duality, the condensate of chargons (the
superconductor) is the vacuum of vortices; the condensate of
vortices is an electronic insulator, where the chargons are
gapped. Within the vortex theory, the superconductor is trivial (being
just the vacuum) and is therefore a good place to plant our feet. From
this vantage, we look out of the superconductor at the neighboring
insulating phases.
The duality transformation, on the lattice, in the presence
of the $Z_2$ gauge field, has been explicitly implemented in Ref.\cite{Z2}.
The full resulting action at half-filling is:
\begin{eqnarray}
S &=& S_s + S_v + S_a + S_{CS} , \\
S_v &=& -t_v \sum_{<ij>} \mu_{ij} cos(\theta_i - \theta_j +
\frac{a_{ij}}{2}) , \\
S_a &=& \frac{\kappa}{8\pi^2} \sum_{\Box} |\Delta \times a_{ij} - 2\pi 
\hat{\tau}|^2,  \\
S_{CS} &=& \sum i \frac{\pi}{4} (1-\prod_{\Box} \sigma)(1-\mu_{ij}) .
\end{eqnarray}
The spinon action $S_s$ is unchanged.
Here, $e^{i\theta_i}$ creates an $\frac{hc}{2e}$ vortex and the flux
of the U(1) gauge field, $a_{ij}$, is the total electrical current. In
particular, a flux of $2\pi$ through a spatial plaquette represents a
charge of $e$.  The terms $S_v$ and $S_a$ together form the usual
dual vortex representation for charge $2e$ Cooper pairs except that
here, the vortices are minimally coupled to the additional ($Z_2$)
gauge field $\mu_{ij}= \pm 1$. The BCS vortex and the spinon are relative
semions; upon circling a vortex, the spinon wavefunction picks up a
minus sign. The term $S_{CS}$ is the $Z_2$ analog of a
Chern-Simons term for the two $Z_2$ gauge fields and mediates this
statistical vortex-spinon interaction.
The spinons ``see'' a $Z_2$ flux $\prod_{\Box} \sigma  =
-1$ attached to each $\frac{hc}{2e}$ vortex, while the vortices see
a flux of $\prod_{\Box} \mu = (-1)^{J_f}$. This flux attachment may be
familiar to many in the
context of the Quantum Hall Effect, where the gauge fields involved
are for the U(1) group. Because of the anomalous ``$ff$'' terms in the
action, spinon number is not conserved, and the usual Chern-Simons term
cannot be used.
 
In the superconducting state, we are in the vacuum of vortices. The
spinons see no flux and are free to propagate independently of the
chargons. However, when single
vortices condense, the long-range statistical interaction
between the BCS vortex and the spinon drives spin-charge
confinement. In the language of Ref.\cite{Z2}, the condensation of
$hc/2e$
vortices is accompanied by a condensation of the visons (vortices in
the Ising field, $\sigma$), leading to a confined phase of
electrons.

We wish here to explore in some detail the
nature of this confinement transition, where the freely-propagating
spin and charge excitations are ``glued together'' to form the
electron. Aspects and implications of this quantum critical point
pertaining to the high $T_c$ phase diagram have been introduced in
Ref. \cite{QCT}. First, we will use Landau theory to find phases
related to the d-wave superconductor by a second-order phase
transition. Then, we will
consider a special case where we recover a
U(1) symmetry for the spinons and will use quantum field theory
methods to
extract some analytic critical properties of the transition
between deconfined and confined phases.

\section{dual vortex theory at half-filling}

Concentrating on the vortices for the time being and working at half-filling,
the dual theory for the charge sector becomes
\begin{equation}
\label{vort}
{\cal L}_{v} = -t_v \cos(\theta_{i} - \theta_{j} + \frac{a_{ij}}{2}) +
\frac{1}{2} \left| \nabla \times \mbox{\bf a} - 2\pi\hat{\tau}\right|^2.
\end{equation}
To obtain a low-energy effective theory, we work with a ``soft-spin'' model
where the vortex creation operator $ e^{i\theta} $ is replaced by a
complex field
$\Phi$. In the interest of exploring the simplest case we set the
charge per unit cell to be exactly e. In the dual theory, this
corresponds to setting 
\begin{equation}
\langle (\nabla \times \mbox{\bf a})_{\hat{\tau}} \rangle = 2 \pi .
\end{equation}
In this section, we drop fluctuations of the gauge field {\bf a}, and
consider a Landau mean-field approach. This is
justified when the on-site repulsion between the electrons, U, is
large. The vortices now 
see exactly $ (\vec{\nabla} \times \vec{\frac{a}{2}})_{\hat{\tau}} =
\pi $ flux per spatial
plaquette, and we are left with the two-dimensional fully-frustrated
quantum XY model: 

\begin{eqnarray}
S = \int d\tau \left \{ \sum_{\vec{r}} |\partial_{\tau} \Phi_{\vec{r}}|^2
- \sum_{\langle\vec{r},\vec{r}'\rangle} t_{\vec{r}\vec{r}'}
(\Phi_{\vec{r}}^* \Phi_{\vec{r}'} + c.c.)  \right. \nonumber \\
\left. + \sum_{\vec{r}}[m^2|\Phi_{\vec{r}}|^2 +
  u(|\Phi_{\vec{r}}|^2)^2]\right\}, 
\end{eqnarray}
where $\vec{r}$ labels sites on the 2d square lattice dual to the
original electron lattice and the sign of $t_{r,r'}$ around a
plaquette is $-1$. The sites of the dual lattice are at the
centers of the plaquettes of the original lattice, and in units of the
lattice constant ($a=1$), $\vec{r} = (x,y)$ with x and y integers. 

\begin{figure}
\epsfxsize=3.5in
\centerline{\epsffile{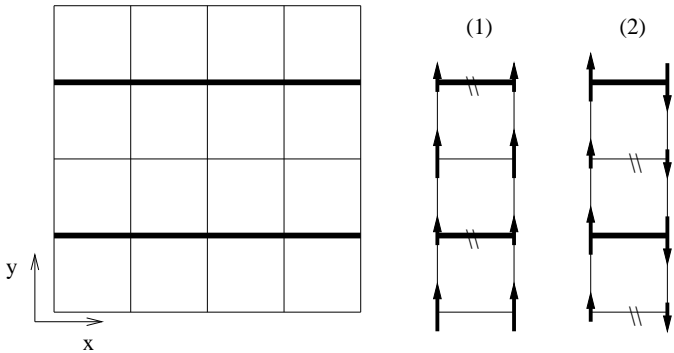}}
\vspace{0.15in}
\caption{ Representation of the fully frustrated 2d lattice. Dark lines
 show the location of negative or ``antiferromagnetic'' bonds
 ($t_{r,r'} < 0$). The two
 low-energy modes $\chi_r^0$ and $\chi_r^{\pi} $ are shown in (1) and
 (2), respectively. Long arrows have length $2+\sqrt{2}$ and short
 arrows have length $\sqrt{2}$. Frustrated or ``unhappy'' bonds are 
 marked with a slash.}
\vspace{0.15in}
\label{GaugeChoice}
\end{figure}

We proceed, following closely the work of others on the fully
frustrated quantum Ising model \cite{BMB}, by choosing the gauge (to
be used in the remainder of this 
paper) seen in Figure \ref{GaugeChoice}. We may diagonalize the
kinetic piece of this
action to find two low-energy modes, living at $(k_x ,k_y) = (0,0)$
and $(\pi,0)$, respectively. In
real space, these (unnormalized) eigenvectors are:
\begin{eqnarray}
\chi^{0}_{\vec{r}} &=& (1+\sqrt{2}) - e^{i\pi y} ,\\
\chi^{\pi}_{\vec{r}} &=& e^{i\pi x}\left[ (1+\sqrt{2}) + e^{i\pi y}
\right] , \\
& & \mbox{(x,y integers).} \nonumber
\end{eqnarray} 

For the purpose of characterizing the low-energy behavior of this
vortex system, we consider fields which are linear combinations of
these two low energy modes,
\begin{equation}
\Phi(\vec{r},\tau) = \Psi_0 (\vec{r},\tau) \chi^{0}_{\vec{r}} + \Psi_{\pi}
(\vec{r},\tau) \chi^{\pi}_{\vec{r}} .
\label{rsfield}
\end{equation}
We now have two complex fields, $\Psi_0 (\vec{r})$ and
$\Psi_{\pi} (\vec{r})$, which describe the low-energy configurations
of our vortex system. The phase transitions of the system can be
explored within Ginzburg-Landau theory. The Ginzburg-Landau
Hamiltonian for the two-vortex system must preserve all the symmetries
of the original lattice Hamiltonian, namely: discrete $\hat{x}$ and
$\hat{y}$ translations, rotations by $\frac{\pi}{2}$, and the vortex
U(1) symmetry ($\Phi \rightarrow e^{i\alpha} \Phi$), as well as
hermiticity. In terms of our two complex vortex fields, these symmetry
transformations take a simpler form when expressed in terms of the fields
\begin{equation}
\phi_1 =\Psi_0 + i\Psi_{\pi} ~,~ \phi_2 = \Psi_0 - i\Psi_{\pi},
\end{equation}
as follows:
\begin{eqnarray}
\label{symmphi1}
T_{\hat{x}}: & \phi_1 & \leftrightarrow \phi_2 ,  \\
T_{\hat{y}}: & \phi_1 & \rightarrow i \phi_2  \nonumber \\
             & \phi_2 & \rightarrow -i \phi_1 , \\
\label{symmphi2}
R_{\frac{\pi}{2}}: & \phi_1 & \rightarrow e^{i\pi /4} \phi_1 \nonumber\\
                   & \phi_2 & \rightarrow e^{-i\pi /4} \phi_2 , \\
\label{U1}
U(1): & \phi_a & \rightarrow e^{i\alpha} \phi_a \;\;\mbox{(for a=1 \& 2)} .
\end{eqnarray}
Allowed terms for the action include 
\begin{eqnarray*}
(I): &(|\phi_1|^2)^n + (|\phi_2|^2)^n,& \\
(II): &{(|\phi_1|^2 |\phi_2|^2 )}^{n},& \\
(III): &{[(\phi_1^* \phi_2)^4 + (\phi_1 \phi_2^*)^4 ]}^{n},&
\end{eqnarray*}
(with arbitrary positive integer, n), and combinations of these terms.
Expanding in powers of the fields, we take as our Landau-Ginzburg
action: 
\begin{eqnarray}
S_{LG} = \int d^2 x d\tau \sum_{a=1,2} \left[ |\partial_{\mu}\phi_a|^2
+ r|\phi_a|^2 \right]  \nonumber \\
+ u_4 (\textstyle{\sum_a} |\phi_a|^2)^2   
+ v_4 |\phi_1|^2 |\phi_2|^2 - v_8 [(\phi_1^*\phi_2)^4 + h.c. ] ,
\label{LG}
\end{eqnarray}
where $\tau$ has been rescaled to set the vortex velocity $\mbox{v}_v
=1$. The terms labeled by $u_4$ and $v_4$ are the only allowed
quartic terms, and are invariant under independent U(1)
transformations on $\phi_1$ and $\phi_2$. We  
have kept the $v_8$ term because it is the lowest-order term which
breaks this symmetry down to the global $U(1)$ of Eqn. \ref{U1}. This
model will be employed to construct a
description of various phases proximate to the d-wave
superconductor within mean field theory. 

We wish to characterize the various states of this vortex
system. It is important to emphasize at this point that not all vortex
condensates destroy superconductivity. Superconductivity is destroyed
when the dual U(1) symmetry of the vortex theory (Eqn. \ref{U1}) is
broken. Therefore, 
it is possible to have non-trivial vortex condensates which are
superconducting. This leads to two scenarios for the
superconductor-insulator transition at half-filling. First, we
may consider
superconductors which are described by a vacuum of vortices;
superconductivity is then destroyed when single vortices
proliferate and condense (in a way which breaks the dual
U(1)). Alternatively, the superconducting state could itself be a
U(1)-preserving vortex condensate which then undergoes a transition
which breaks the dual U(1), killing superconductivity. 

In the following sections, we explore the phases of our dual vortex
model using the Landau-Ginzburg action of Eqn. \ref{LG}. Due to the
frustration of the vortex theory with one electron per unit cell, the
vortex condensates will break lattice symmetries. Some of these
spatially-ordered states are superconductors and some are
insulators. We will begin by describing the possible superconducting
states within the dual theory (including, a \emph{striped}
superconductor), and then move on to a description of 
the insulating states. 
Ignoring charge fluctuations in the superconducting states (as we have in
arriving at Eqn. \ref{LG}) is not justified, and a good description of
these states would require putting the charge fluctuations back
in. Here, we content ourselves to characterizing the phases of our
vortex system by their broken symmetries. We conclude with a summary
of the possible 
transitions from superconductor to insulator within this mean field theory.
\subsection{Superconductors}

\subsubsection{Vortex Vacuums} 

The simplest superconducting phase is just the vortex vacuum. This is
the standard BCS d-wave superconductor. Destruction of
superconductivity occurs when single vortices proliferate out of the
vacuum and condense, breaking the dual U(1) symmetry. 
The effective action for this transition is Eqn. \ref{LG}.

\subsubsection{Paired Vortex Condensates}

Condensation of single $hc/2e$ vortices necessarily breaks the dual
U(1) symmetry (Eqn. \ref{U1}), killing superconductivity. However,
when \emph{pairs} of vortices condense, the U(1) can be
preserved. Consider the paired vortex condensate:
\begin{equation}
\langle \phi_2^*\phi_1 \rangle \neq 0,\; \langle \phi_1 \rangle =
\langle \phi_2 \rangle = 0 . 
\end{equation}
We see that in this condensate, the dual U(1) is preserved, and
the state is characterized by the phase of the condensate (setting the
amplitude $|\langle \phi_2^*\phi_1 \rangle| = 1$ for simplicity),
\begin{eqnarray}
\langle \phi_2^*\phi_1 \rangle &=& e^{i\theta}, \\
\theta &\equiv& \theta_1(x) - \theta_2(x).
\end{eqnarray}
Here, $\theta_1$ and $\theta_2$ are the phases of $\phi_1$ and
$\phi_2$, respectively, and are still free to fluctuate. Only the
combination $\theta = \theta_1 - \theta_2$ is uniform, 
reflecting the fact that the dual U(1) symmetry is preserved (i.e.,
$\phi_1$ and $\phi_2$ are uncondensed). 
The only term in the Landau-Ginzburg action which depends on $\theta$
is the $v_8$ term, giving:
\begin{equation}
S_v = -v_8 \int d^2xd\tau \cos(4\theta) .
\end{equation}
We see that the ground state depends on the sign of $v_8$:
\begin{eqnarray}
v_8>0 &:& \theta = n \frac{\pi}{2} ,\\
v_8<0 &:& \theta = \frac{\pi}{4} + n \frac{\pi}{2} , 
\end{eqnarray}  
with n an integer.

The spatial symmetries in Eqns.\ref{symmphi1}-\ref{symmphi2}, written
in terms of the relative phase $\theta$, are: 
\begin{eqnarray}
\label{symmtheta1}
T_{\hat{x}}:  \theta & \rightarrow & - \theta ,\\
T_{\hat{y}}:  \theta & \rightarrow & \pi - \theta ,\\
R_{\frac{\pi}{2}}:  \theta & \rightarrow & \theta + \frac{\pi}{2}.
\label{symmtheta2}
\end{eqnarray}
From this we can see that the vortex condensate favored by $v_8 >0$
breaks the lattice rotational symmetry and \emph{one} of the two
translational symmetries. We therefore associate this condensate with
a stripe-type ordering: a \emph{striped superconductor}. This state is
particularly interesting given recent experimental results which
suggest possible stripes in the superconducting state of $La_{2-x}Sr_x
CuO_4$ \cite{stripedSC}.
The ground
state for $v_8<0$ breaks all of the lattice symmetries; we identify
this state with a ``plaquette'' order which will be made more explicit in
upcoming sections when we discuss the insulating states of the vortex
system. For now, we emphasize the possibility of spatially-ordered
superconducting states which emerge quite naturally within our dual
vortex description.

Still working in the dual description, these striped and plaquette
superconductors are described by an effective theory of \emph{one}
vortex species, since the paired condensation has locked the two
original vortices together: the vortex phases $\theta_1(x)$ and
$\theta_2(x)= \theta_1(x) - \theta$ still fluctuate within the
superconducting phase, but not independently. When the remaining phase
$\theta_1$ becomes constant over the sample, the dual U(1) is
broken, and superconductivity is destroyed.
Therefore, for these spatially ordered superconductors, $S_{LG}$
(Eqn. \ref{LG}) reduces to:
\begin{equation}
\label{ssc}
S_v = \int d^2x d\tau \left[ |\partial_{\mu}\phi_1|^2 + r|\phi_1|^2 + u
(|\phi_1|^2)^2 \right] .
\end{equation} 
It is worth noting that we have gone from a theory of a single
fully-frustrated vortex to a theory of a single unfrustrated
vortex via a theory of two vortices. This is possible because in a
striped or plaquette  
superconductor, the unit cell is doubled. If one started from
scratch in constructing a dual theory of these striped (plaquette)
superconductors, the vortices would see
$2\pi$ rather than $\pi$ flux per (doubled) unit cell and there would
be only one low-energy mode. 

\subsection{Confined Insulators}

When single vortices condense at half-filling, we move from the
d-wave superconductor into a confined insulator.
Within our dual formulation, these insulators are described by
condensates which break the dual U(1) symmetry of Eqn. \ref{U1}. 
In the case of superconductors which are vortex vacuums, 
because we have {\em two} vortex species, there are many
ways to do this and therefore many possible single vortex condensates.  
We will see that these different vortex condensates
correspond to different insulating states of electrons.
We return to the case of the striped and plaquette superconductors after
first enumerating the insulating states at the mean field level,
using the action of Eqn. \ref{LG}.

The most general U(1)-breaking vortex condensate is:
\begin{eqnarray}
\langle \phi_1 \rangle &=& |\langle \phi_1 \rangle| e^{i\theta_1}, \\
\langle \phi_2 \rangle &=& |\langle \phi_2 \rangle| e^{i\theta_2}, 
\end{eqnarray}
where $|\langle \phi_1 \rangle|$, $|\langle \phi_2 \rangle|$,
$\theta_1$, and $\theta_2$ are all fixed real numbers.
Within our dual Landau-Ginzburg model, condensing the vortices
corresponds to setting $r<0$ and $u_4 >0$. The signs of $v_4$ and
$v_8$ then determine the ground state. For $v_4 <0$, both vortex
species acquire a non-zero amplitude $|\langle \phi_1 \rangle| = |\langle
\phi_2 \rangle| \neq 0$ and their relative phase $\theta_{12}
=\theta_1 - \theta_2$ is determined by the sign of $v_8$. On the
other hand, if $v_4>0$, the ground states are condensates of either
$\phi_1$ {\em or} $\phi_2$ and the sign of $v_8$ is irrelevant. Each
of these condensates will correspond to a different insulating state
of the electron system.  We consider each case in turn.

\subsubsection{ $|\langle \phi_1 \rangle| = |\langle
\phi_2 \rangle| \neq 0  $ }

These condensates are favored by $v_4 <0$, and the relative phase
($\theta_1 - \theta_2$) is determined by the sign of $v_8$. Taking the
magnitudes $|\langle \phi_1 \rangle| = |\langle
\phi_2 \rangle| = 1$, this term in
the action can be rewritten as:
\begin{equation}
- v_8 [(\phi_1^*\phi_2)^4 + h.c. ] = - v_8 \cos (4\theta_{12}).
\end{equation} 
In terms of this relative phase, the spatial symmetries
are given by Eqns. \ref{symmtheta1}-\ref{symmtheta2} with the
replacement $\theta \rightarrow \theta_{12}$.

\subsubsection*{ \protect \flushleft \underline{$ \theta_1 - \theta_2
= n \frac{\pi}{2} $} }

\begin{figure}
\epsfxsize=3.5in
\centerline{\epsffile{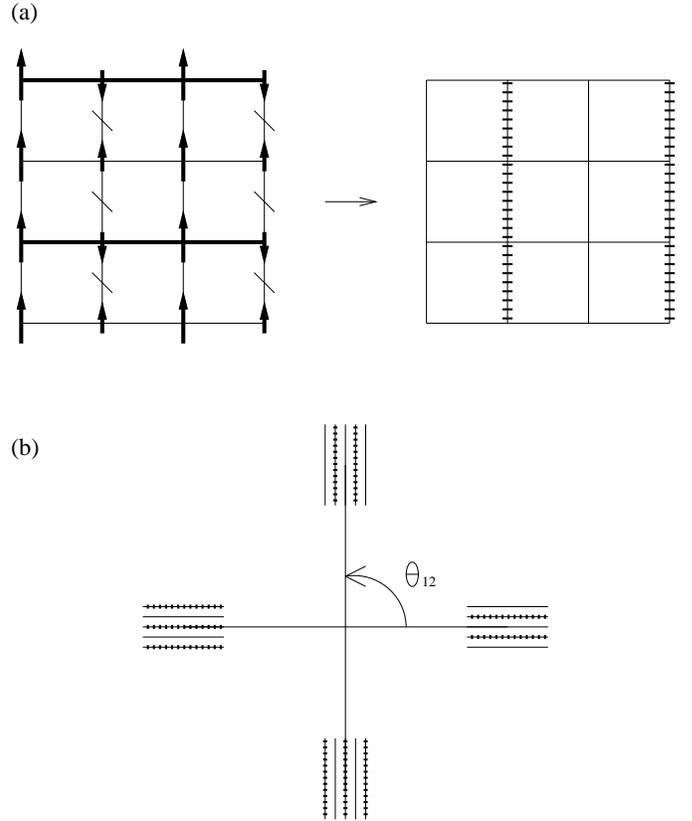}}
\vspace{0.15in}
\caption{(a)The vortex condensates with $\theta_{12} = \frac{\pi}{2}$,
which correspond to $\Phi(r) = \frac{1}{\sqrt{2}} (\chi^0_r +
\chi^{\pi}_r )$. Frustrated bonds are slashed.
(b) Schematic of the four ``striped'' states corresponding to 
$\theta_{12}= n \frac{\pi}{2} $ with higher energy (frustrated) bonds 
slashed.}
\vspace{0.15in}
\label{thetapi2}
\end{figure}

This class of condensates is preferred by $v_8 >0$. 
There are four general states, corresponding to each of the possible
values of $n$. We see by the symmetry transformations
in Eqns.\ref{symmtheta1}-\ref{symmtheta2} that each of these states 
breaks the lattice rotational
symmetry as well as breaking {\em one} of the two translational
symmetries while leaving the other intact. On these grounds alone, we
could guess that these states correspond to ``stripe-like'' phases. To
be more concrete, we may go back to our real-space representation for
the vortex field $\Phi (\vec{r})$ in Eqn.\ref{rsfield}  and draw real-space
pictures of these lattice states. The values of the fields at various 
points will be gauge dependent, but the location of frustrated bonds (which
are places of higher energy density) is gauge-independent
and therefore a good way to characterize the state of the system. This
is shown for the case $\theta_{12} = \frac{\pi}{2}$, as an example, 
in Figure \ref{thetapi2}a. Investigations of this sort lead us to
conclude that the
four ground states of the system in this case are characterized by
``stripes'' of energy-density as shown in Figure \ref{thetapi2}b . 

\begin{figure}
\epsfxsize=2.0in
\centerline{\epsffile{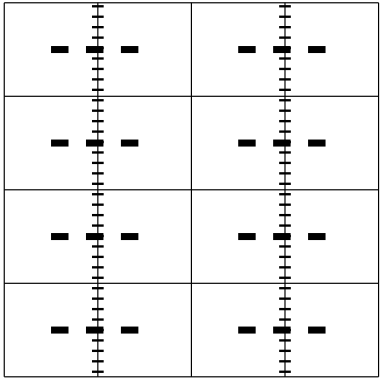}}
\vspace{0.15in}
\caption{Schematic of the relationship between frustrated bonds of
 the dual lattice (slashed) and links of the original lattice where the
 corresponding ``singlet bonds'' live (dashed).} 
\vspace{0.15in}
\label{v8g}
\end{figure}

We now turn to a characterization of this system in terms of the
electron degrees of freedom. Because we have broken the dual U(1)
symmetry of the vortices and we are at half-filling, these states
will be Mott insulators.
The charge degrees of freedom live on the plaquettes of the dual
lattice and are fixed at one charge of $e$ per dual plaquette. It has
been suggested \cite{SV,Z2} that the frustrated bonds of the dual
lattice should correspond to singlet bonds of the electron system,
since one expects regions of higher energy along the links 
where the electrons spend most of their time. This relationship
between the frustrated bonds on the dual lattice and the singlet
bonds on the original lattice is illustrated in Figure \ref{v8g}. These
``striped'' vortex phases then correspond to spin-Peierls (or ``bond 
density wave'') order in the insulating electron system.

\subsubsection*{ \protect \flushleft \underline{$\theta_1 - \theta_2 =
 \frac{\pi}{4} + n\frac{\pi}{2} $} }

\begin{figure}
\epsfxsize=3.5in
\centerline{\epsffile{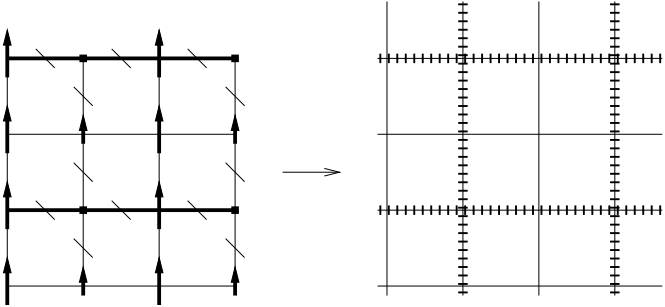}}
\vspace{0.15in}
\caption{The vortex condensate with $\theta_{12} = \frac{\pi}{4}$,
which corresponds to $\Phi_r = 0.9 \chi_r^0 + 0.4
\chi_r^{\pi}$. Locations where the field is zero are denoted by a dot.
Relatively unhappy bonds are slashed.}
\vspace{0.15in}
\label{v8l}
\end{figure}

These condensates are favored by $v_8<0$. 
Here, however, each ground state breaks all of the discrete lattice
symmetries. We proceed as above and obtain characterizations of these
states in terms of the location of frustrated bonds. The result is
a plaquette-like structure, as seen in Figure \ref{v8l}.

\begin{figure}
\epsfxsize=2.0in
\centerline{\epsffile{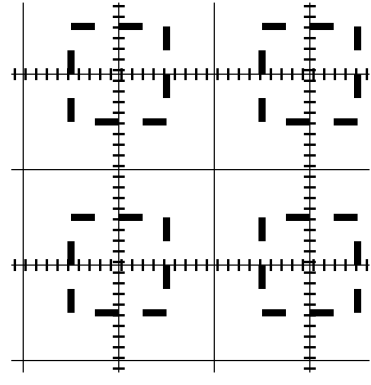}}
\vspace{0.15in}
\caption{Schematic of states corresponding to the $v_8 < 0 $ vortex
  condensates. Relatively unhappy bonds of the dual lattice are 
  slashed, and links of the original lattice where the singlet-type bonds 
  live are dashed. We 
  associate this structure with a ``plaquette RVB'' state of 
  the electrons.}
\vspace{0.15in}
\label{PlaqRVB}
\end{figure}

In terms of the electron degrees of freedom, we would like to again
interpret the frustrated bonds of the dual lattice as regions where
singlet-type bonds of the electron system reside. The plaquette-like
structure of these vortex states may then correspond to a ``plaquette RVB''
state of the electron system, as shown in Figure
\ref{PlaqRVB}.

\subsubsection{ $\langle \phi_1 \rangle \neq 0, \langle
  \phi_2 \rangle = 0 $ or $\langle \phi_1 \rangle = 0,\langle \phi_2 \rangle
\neq 0 $ }

\begin{figure}
\epsfxsize=3.5in
\centerline{\epsffile{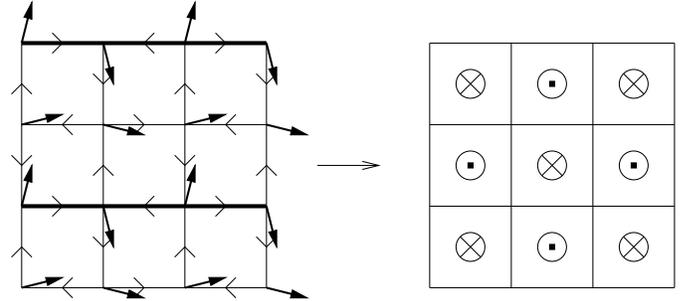}}
\vspace{0.15in}
\caption{The vortex condensate with $\Phi_r = \chi_r^0 +
i\chi_r^{\pi}$. Since the field is complex, the magnitude is given by
the length and the argument by the direction of the arrow at
each site. The direction of vortex current is indicated on the bonds.}
\vspace{0.15in}
\label{v4g}
\end{figure}

These condensates are preferred in the case $v_4 > 0$. 
We may proceed as above in drawing real-space diagrams corresponding
to these states. We find, as shown in Figure \ref{v4g}, that these states
have vortex currents around each plaquette, of alternating sign.

In order to interpret this state, we will have to put back in the
spinons which have been ignored in the previous discussion. The
vortex-spinon action is: 
\begin{eqnarray}
S &=& S_v + S_s + S_{CS}, \\
S_v &=& -t_v \sum_{<i'j'>} \mu_{i'j'} \cos(\theta_{i'} - \theta_{j'} - 
\frac{a_{i'j'}}{2}), \\
\mbox{with} & & \langle \nabla \times \mbox{\bf a}\rangle = 2 \pi, \\
S_s &=& -\sum_{<ij>} \sigma_{ij} [t^s_{ij} \bar{f}_i f_j +
t^{\Delta}_{ij}f_{i\uparrow}f_{j\downarrow}] - \sum_i \bar{f}_i f_i
,\\
S_{CS} &=& \sum i\frac{\pi}{4}(1-\prod_{\Box'}\mu)(1-\sigma_{ij}), 
\end{eqnarray}
where $i,j$ label sites on the original lattice and $i',j'$ label
sites on the dual lattice.
Looking at $S_v$, we see that the alternating vortex currents would
like to induce compensating fluctuations in either the $a_{i'j'}$ or 
$\mu_{i'j'}$ fields. Allowing
fluctuations of the gauge field {\bf a} (which describes charge
fluctuations), and ignoring the coupling to the spinons, the
alternating vortex currents would induce CDW order 
at wavevector $(\pi,\pi)$. However, with a large on-site U, this state
will be greatly suppressed. 
If we forbid charge fluctuations, we see that the alternating vortex
currents will instead drive 
a mean field in the $Z_2$ gauge field:
\begin{equation}
\prod _{\Box'} \mu_{i'j'} = (-1)^{n_f} \approx -1,
\end{equation}
(where $n_f$ is the number of spinons in the dual plaquette denoted by
$\Box$), which corresponds to one spinon per unit cell. Unlike the
previously-considered vortex condensates (with $v_4 < 0$), at the
level of vortex mean field theory, this state has no broken
translational symmetries. (However, we cannot rule out the breaking of     
symmetries by the charge and spin fluctuations.)
We note that a possible candidate for this state
which has one electron per until cell and uniform energy density
is the antiferromagnet.  

\subsection{Summary of Vortex Theory}

We have seen that our dual vortex theory describes both standard BCS
and striped or plaquette d-wave superconductors as well as a host
of confined insulating states. We now summarize the results of Landau
theory for the transitions from the d-wave superconductor to the
confined insulator at half-filling.

We consider first the transition from a vacuum of
vortices (a superconductor) to a 
U(1)-breaking condensate of vortices (an insulator). 
Within mean field theory, the nature 
of the insulating state is determined by the signs of the coupling
constants in Eqn \ref{LG}, and we have the following possible
\emph{direct} transitions out of the symmetric d-wave
superconductor:
\begin{eqnarray}
\label{states1} 
&v_4<0,\; v_8>0;&\; dSC \rightarrow \mbox{spin-Peierls,} \\
&v_4<0,\; v_8<0;&\; dSC \rightarrow \mbox{plaquette RVB,} \\
&v_4>0;&\; dSC \rightarrow \mbox{uniform state of electrons}.
\label{states2} 
\end{eqnarray}

One might hope to ascertain which of these insulating states is
preferred close to a d-wave superconductor, including
fluctuations beyond the mean field level, by considering the
fixed points of the action in Eqn.\ref{LG}. 
In particular, we see that the sign of $v_4$ determines
whether we enter one of the states of broken translational symmetry
(spin-Peierls or plaquette RVB), or the state with uniform energy-density
(possibly the antiferromagnet).
The work of Blagoeva \cite{EB} on the theory of
two-component complex fields with these (and other) couplings gives
a stable fixed point at $v_4 <0$, to order $\epsilon ^2$ ($\epsilon =
4-D$, $D=d+1$, in d spatial dimensions). This
suggests that the transition dSC $\rightarrow$ 
spin-Peierls 
would be preferred over dSC $\rightarrow$ uniform state. This is tantalizing
given the experimental evidence for intervening ``stripey''
phases between the 
superconducting and antiferromagnetic phases in the cuprates
\cite{stripedPG}. 

In the case of the striped and plaquette superconductors, 
when the single vortex species in Eqn. \ref{ssc} condenses,
superconductivity in these states is destroyed and we enter a confined
insulating state. Because the relative phase $\theta_1 - \theta_2$ is
already fixed within these superconductors, we see from our above
analysis of the insulating phases that the insulating state is
pre-determined. The striped 
superconductor (with $\theta = n\pi/2$) enters the spin-Peierls
insulator, and the plaquette superconductor (with $\theta = \pi/4 +
n\pi/2$) enters the plaquette-RVB insulator. In other words, these
spatially-ordered superconductors make transitions into insulating
states with the same broken spatial symmetries:
\begin{eqnarray}
\label{stripes1}
v_8 >0;\; \mbox{striped SC} &\rightarrow& \mbox{spin-Peierls,} \\
v_8 <0;\; \mbox{plaquette SC} &\rightarrow& \mbox{plaquette-RVB.}
\label{stripes2}
\end{eqnarray}

In the preceding section, we have considered states of electron
systems at half-filling near a d-wave superconductor within a dual
formulation in terms of
vortices. Each phase is characterized by a dual (vortex) order
parameter. At one electron per unit cell, the frustration of the
vortex theory manifests itself in spontaneously broken spatial
symmetries. Exploiting the fact that the vortex order parameters break
spatial symmetries has helped us identify these vortex phases with
more familiar phases of electrons (such as the spin-Peierls state), as
well as phases like the striped superconductor. The dual formulation
shows us the enhanced chance for striped superconductors near
half-filling. 
In the next section, we will add back in the spinons
(and along with them, their 
long-ranged statistical interaction with the vortices), and 
extract information about the critical properties of the confinement
transition using field theory methods. 

\section{confinement transition}

Having explored the vortex sector of the theory with one electron per
unit cell, we
now wish to put the spinons back in and address the critical
properties of the
confinement transition. Because we will continue to work at
half-filling, the confined states of electrons will be Mott
insulators. 
While the theory of vortices and spinons coupled to $Z_2$ gauge
fields may in principle be numerically accessible, the action
suffers
from the notorious fermion sign problem. Here, we discuss a special
case which will turn out to be accessible
to perturbative RG calculations.

Focusing on the spinon Hamiltonian (and dropping the $Z_2$ gauge field
for the time being):
\begin{equation}
H_s = - \sum_{<rr'>} [t_{rr'}^s f^{\dag}_{r}f_{r'} + 
t^{\Delta}_{rr'}(f_{r\uparrow}f_{r'\downarrow}+c.c.)] ,
\label{spinons}
\end{equation}
we choose the special case:
\begin{equation}
t^s = |t^{\Delta}| \equiv t,
\end{equation}
with nearest-neighbor d-wave pairing amplitude:
\begin{equation}
\begin{array}{cc} t^{\Delta}_{\vec{r},\vec{r}\pm \hat{x}} & =  +t ,\\
                         t^{\Delta}_{\vec{r},\vec{r}\pm \hat{y}} & =
                         -t. \\ \end{array}  
\end{equation}
Following Affleck \cite{IA}, we introduce the fields 
\begin{equation}
\left( \begin{array}{c} d_{r\uparrow} \\ d^{\dag}_{r\downarrow}
  \end{array} \right)
= \left\{ \begin{array}{ll} e^{-i\frac{\pi}{8} \sigma_y} \left( 
\begin{array}{c} f_{r\uparrow} \\ f^{\dag}_{r\downarrow} \end{array}
\right), & \mbox{ for $y$ even, } \\
(-i\sigma_y) e^{-i\frac{\pi}{8} \sigma_y} \left( 
\begin{array}{c} f_{r\uparrow} \\ f^{\dag}_{r\downarrow} \end{array} 
\right), & \mbox{for $y$ odd,} \end{array} \right.
\end{equation}
(where $\sigma_y$ is the usual Pauli matrix), the spinon Hamiltonian
becomes:
\begin{equation}
H_s = -\sum_{<rr'>} t_{rr'} (d^{\dag}_{r\alpha} d_{r'\alpha} + h.c.) ,
\end{equation}
with
\begin{equation}
  t_{rr'} = \left\{ \begin{array}{ll} -t & 
\mbox{for $y$ and $y'$ even,} \\
                            t & \mbox{else.} \end{array} \right.  
\end{equation}
This is the Hamiltonian of fermions hopping in 2d in the presence of
$\pi$ flux per plaquette. We have succeeded in finding a Hamiltonian 
for the spin sector which has a conserved fermion number. 
The original theory (Eqn.\ref{ElecTheory})
can now be written in terms of these $d$ fermion fields, the
chargons and the $Z_2$ gauge field. Following a transformation which
can get rid of the Berry's phase term \cite{Berry} this theory can be
modeled numerically with no fermion sign problem. 
Here, we instead proceed to  
a low-energy continuum Hamiltonian for the spin sector.
To this end, we diagonalize $H_s$ to find two
Dirac points. These are the usual d-wave quasiparticle nodes at 
$(k_x,k_y)=(\pm \frac{\pi}{2},\pm \frac{\pi}{2})$ except that due 
to the $\pi$
flux per plaquette, we have doubled the unit cell and halved the
Brillouin zone; it now contains only two of these nodes, which we
denote $\vec{K}_1$ and $\vec{K}_2$.
In terms of long-wavelength fields living at these two nodes,
\begin{equation}
d_{j \alpha}(\vec{x}) \simeq \psi_{1j\alpha}(\vec{x}) e^{iK_1\cdot x} +
\psi_{2j\alpha}(\vec{x})  
e^{iK_2\cdot x},
\end{equation}
(where $j=1,2$ labels the sublattice), the continuum Hamiltonian is:
\begin{eqnarray}
H_s &=& \int d^2 x \;\mbox{v}_s \psi^{\dag}_{1\alpha}[\tau_1
(-i\partial_x) + \tau_2(-i \partial_y)]\psi_{1\alpha} \nonumber \\
  & & + \mbox{v}_s \psi^{\dag}_{2\alpha}[\tau_2 (-i\partial_x) +
\tau_1(-i \partial_y)]\psi_{2\alpha},
\end{eqnarray}
where,
\begin{eqnarray}
\tau_1 &\equiv & \frac{1}{\sqrt{2}}(\tau_x + \tau_z) ,\\
\tau_2 &\equiv & \frac{1}{\sqrt{2}}(\tau_x - \tau_z) .
\end{eqnarray}
Here, $\vec{\tau} $ acts in the sublattice space,
and we have rotated the x and y axes at each node by
$45^{\circ}$.  
Just as the Hamiltonian for the $d$ fermions was diagonal in
the spin label, so is this one, and we are left with a theory of four
species of Dirac fermions. Note that the spinon characteristic velocity,
$\mbox{v}_s$ is isotropic in space because of our choice $t_s =
|t_{\Delta}|$. 

Defining Dirac matrices in 2+1 dimensions:
\begin{equation}
\begin{array}{cc} \mbox{at node}\; \vec{K}_1:\; & \mbox{at node}\;
\vec{K}_2:\; \\
                \gamma_0 \equiv \tau_y,   & \gamma_0 \equiv \tau_y,\\
                \gamma_1 \equiv \tau_2,   & \gamma_1 \equiv -\tau_1,\\
                \gamma_2 \equiv -\tau_1,   & \gamma_2 \equiv \tau_2,\\
\end{array}
\end{equation} 
\begin{equation}
(\gamma_{\mu})^{\dag} = \gamma_{\mu} \;;\;  \{ \gamma_{\mu} , \gamma_{\nu}
\} = 2\delta_{\mu \nu} \; \mbox{(at each node)}, 
\end{equation}
we proceed to the Euclidean Lagrangian density:
\begin{eqnarray}
{\cal L}_s &=& \overline{\psi}_a [\gamma_0 \partial_0 + \mbox{v}_s
 \gamma_i \partial_i]\psi_a , \\
\overline{\psi} &\equiv& \psi^{\dag} \gamma_0 . 
\end{eqnarray}
The fields $\psi_a $ have two components (corresponding to the
sublattice label), and summation conventions on the number of species
$a \in [1,4]$ (one for each spin at each of the two nodes) and the
spatial dimension $i \in [1,2]$ are in use.

We have succeeded in writing a low-energy effective theory for the 
spin sector which is just that of four species of
two-component Dirac fermions. We may now write down a full low energy
effective theory where, since the spinon and vortex sectors each
display U(1) symmetries:
\begin{eqnarray}
\psi_{a} &\rightarrow & e^{i\alpha_{\psi }} \psi_{a}, \\
\phi &\rightarrow & e^{i\alpha_{\phi }} \phi,
\end{eqnarray}
we may implement the statistical spinon-vortex interaction using U(1) 
(rather than $Z_2$) Chern-Simons gauge fields, $\mbox{{\bf A}}^{\phi}$
and $\mbox{{\bf A}}^{\psi}$. 
We proceed to a field theory modeling the
confinement transition between a spin-charge separated (d-wave)
superconductor and a spin-charge confined Mott insulator. For
simplicity, we consider the vortex theory with only one species
(Eqn.\ref{ssc}), and thereby consider transitions out of the
striped/plaquette superconductor given in Eqns. \ref{stripes1} -
\ref{stripes2}. The low-energy effective action is:
\begin{eqnarray}
S &=& \int d^2 x d\tau [{\cal L}_s + {\cal L}_v + {\cal L}_{CS} +
{\cal L}_{vs}], \\ 
{\cal L}_s &=& \overline{\psi}_{a} (\not\!\partial - i
g \not\!\!A^{\psi} )\psi_{a} + \kappa \overline{\psi}_{a} ( \gamma_i
\partial_i - i g \gamma_i A_i^{\psi} )\psi_{a},\\
\label{XY}
{\cal L} _v &=& |(\partial_{\mu} - i g A_{\mu}^{\phi} |^2   
 + m^2 |\phi|^2 + u_0 (|\phi|^2)^2 ,\\
\cal{L}_{CS} &=& i \epsilon_{\mu \nu \lambda} A_{\mu}^{\phi} \partial_{\nu}
 A_{\lambda}^{\psi} ,\\
{\cal L}_{vs} &=& v_0 \psi_{a}^{\dag} \psi_{a} |\phi|^2 ,\\
\mbox{with} & & a \in [1,N=4] \;,\; 
\kappa = \frac{\mbox{v}_s}{\mbox{v}_v} - 1,   
\end{eqnarray}
where $\kappa$ is a measure of the velocity anisotropy between
vortices and spinons and will be treated as a perturbation.
We have added the term ${\cal L}_{vs}$ in the interest of including all
possible relevant interactions.
The Chern-Simons term causes a vortex taken around a spinon to acquire
a phase of:
\begin{equation}
\phi \rightarrow e^{i g \oint \vec{A}^{\phi} \cdot d\vec{l} }\phi  =
e^{ig^2}\phi ,
\end{equation}
and likewise for a spinon after encircling a vortex:
\begin{equation}
\psi \rightarrow e^{i g \oint \vec{A}^{\psi} \cdot d\vec{l} }\psi =
e^{ig^2} \psi ,
\end{equation}
so that the full statistical interaction is achieved when:
\begin{equation}
g^2 = \pi = 2\pi \alpha ; \; \; \alpha = \frac{1}{2}
\end{equation}
(where $\alpha$ is the so-called ``statistics angle'' and is equal to
$1/2$ since the vortex and the spinon are relative semions).
The theory as written neglects charge fluctuations, which is not
justified within the superconducting phase. The full vortex theory
would include an additional minimal coupling to a 
gauge field {\bf a} \cite{fluc}. As seen in the dual XY model, this
coupling causes 
runaway flows, and is probably best modelled numerically. At this point,
we leave out the gauge field {\bf a} and its attendant problems, but we
will revisit this question shortly.

When the vortex Lagrangian is taken through criticality ($m^2 < 0$),
the statistical interaction, mediated by the gauge fields $A_{\mu}^{\phi}$
and $A_{\mu}^{\psi}$, will drive spin-charge confinement. Here, we seek the
effect of these statistics on critical properties of the system. In
particular, we wish to calculate $\beta$ functions for the couplings
$u_0$, $v_0$, $\kappa$, and $g$, as well as the anomalous dimensions
of the vortex and spinon fields.  

We work in $D=d+1=3$ dimensions (indeed, our Chern-Simons flux
attachment is not well-defined in higher dimensions), and define
dimensionless couplings: 
\begin{eqnarray}
u &=& \Lambda^{-1} u_0 K_{D=3} ,\\
v &=& v_0 K_{D=3} ,\\
\end{eqnarray}
where factors of $K_D = {[2^{D-1} \pi ^{D/2} \Gamma(D/2) ]}^{-1}$ have
been put in for later convenience. 
The bare propagators in the Landau gauge are:
\begin{eqnarray}
\mbox{fermions:}\;\; &G^{\psi}_0 & = - \frac{i \not\!k}{k^2} ,\\
\mbox{vortices:}\;\; &G^{\phi}_0 & = \frac{1}{k^2} ,\\ 
\mbox{gauge fields:}\;\; &S^{\mu \nu}_0 & = - \frac{\epsilon^{\mu \nu
    \lambda} k^{\lambda}}{k^2} = \langle A_{\mu}^{\psi} A_{\nu}^{\phi}
\rangle ,\\ 
        & & \langle A^{\phi} A^{\phi} \rangle = \langle A^{\psi}
        A^{\psi} \rangle = 0. 
\end{eqnarray}
(The fermion propagator is diagonal in the label $a$, so we have
suppressed this index.)

For the $\beta$-functions we find, to lowest non-vanishing order (1
loop):
\begin{eqnarray}
\frac{du}{dl} &=& u - 10 u^2 + (\frac{N}{3} + C\kappa ) v^2 + \cdots ,\\ 
\frac{dv}{dl} &=& - 4 uv + \cdots ,\\
\frac{dg^2}{dl} &=& 0 ,\\
\frac{d\kappa}{dl} &=& 0 + \cdots . 
\end{eqnarray}
We expect that at higher orders, $g$ will enter into $\frac{du}{dl}$ 
and $\frac{dv}{dl}$ non-trivially, but that $g$ itself should not 
renormalize at any order, following the argument given by Semenoff 
{\em et al.} \cite{SSW}. 
The one-loop RG equations for $u$ and $v$ have a stable
solution at $v=0$, so that the theory decouples into separate spinon
and vortex theories.  
At this order, since the
spinon and vortex sectors decouple, we may ignore the Chern-Simons
gauge fields (effectively taking $g=0$) and include the effects
of charge fluctuations by using the full dual 
XY model for the vortex sector:
\begin{eqnarray}
{\cal L} _v &=& |(\partial_{\mu} - i e_0 a^{\mu})\phi |^2 +
\frac{1}{2}|\vec{\nabla} \times \vec{a}|^2 \nonumber \\  
& & + m^2 |\phi|^2 + u_0 (|\phi|^2)^2 .
\end{eqnarray}
Recently, much work has gone into tackling the critical properties of
the $e\neq 0$ model \cite{HS}, and we may use these results.

To first order, then, we find a \emph{fixed line}, parameterized by
values of the statistics angle 
$\alpha$ (or equivalently, the coupling $g$). At lowest non-vanishing
order, this line is given by:
\begin{eqnarray}
u^* &\simeq& u^*_{dual} ,\\
v^* &\simeq& 0 ,\\
(g^2)^* &=& g^2 = \pi , \\
e^* &\simeq& e_{dual} , \\
\kappa^* &\simeq& \kappa ,
\end{eqnarray}
where, by $u^*_{dual}$ and $e_{dual}$, we mean the values of these
couplings at the fixed point of the dual XY model.

In order to see whether spinon-vortex
velocity anisotropy grows, we need to take the $\beta$ function for
$\kappa$ to its lowest non-vanishing order, which is two loops. The
result is: 
\begin{equation}
\frac{d\kappa}{dl} = - \frac{31}{240} \frac{g^4}{\pi^2} \kappa + \cdots .
\end{equation}
Since the system flows toward $\kappa = 0$, it is legitimate to treat
this term as a perturbation, and the theory becomes ``relativistic'' at
the critical point.

We proceed by calculating the anomalous dimensions of the
spinon and vortex fields, to lowest order, near the critical point. 
To that end, we consider the self-energies:
\begin{eqnarray}
{[G^{\phi}(k)]}^{-1} &=& {[G_0^{\phi} (k)]}^{-1} + \Sigma^{\phi}(k), \\
{[G^{\psi}(k)]}^{-1} &=& {[G_0^{\psi}(k)]}^{-1} +\Sigma^{\psi}(k). \\
\end{eqnarray} 
Near the critical point, the anomalous dimensions are given by:
\begin{eqnarray}
G^{\phi}(k) &\propto& \frac{1}{|k|^{2- \eta_{\phi}}} , \\
G^{\psi}(k) &\propto& \frac{-i \not\! k}{|k|^{2- \eta_{\psi}} } 
\end{eqnarray} 
(up to additive constants). Working at the fixed point $(u,v,g^2) =
(u^*=u^*_{dual}, v^*=0, {g^2}^* =\pi)$ and
calculating the spinon and vortex self-energies to two loops 
in 3 dimensions, we find:
\begin{eqnarray}
\eta_{\phi} &=& \eta_{dual} - \frac{4}{3} \frac{(g^4)^*}{16\pi ^2} N +
\cdots ,\\ 
\eta_{\psi} &=& - \frac{1}{3} \frac{(g^4)^*}{16 \pi ^2} + \cdots . 
\end{eqnarray}

Since we are in the case with one vortex species, we may take the
numerical results of Hove and Sudb\o \cite{HS} 
for the anomalous dimension of the vortex field in the {\em
full} dual XY model in $D=3$: $\eta_{\phi} \simeq -0.24$.
After plugging in $N=4 $, and ${g^2}^* = \pi$ in our result, we find:
\begin{eqnarray}
\eta_{\phi} \simeq -0.24 - \frac{1}{3} \simeq -0.57 ,\\
\eta_{\psi} \simeq -\frac{1}{48} \simeq -0.02 .
\end{eqnarray}

These critical exponents may reveal themselves in many quantities. In
particular, the spectral function as probed by ARPES and the 
spin-spin correlations probed by  NMR or neutron scattering. 
Within our theory, the low-energy electron correlator decouples
into chargon and spinon pieces for $g\rightarrow 0$:
\begin{equation}
\langle c(x) c^{\dag}(0) \rangle = \langle b(x) b^{\dag} (0) \rangle
\langle f(x) f^{\dag}(0) \rangle .
\end{equation}
These correlators will exhibit anomalous dimensions $\eta_{b}$ and
$\eta_{f}$, which can be expanded perturbatively around $g^* = 0$:
\begin{eqnarray}
\eta_b &=& \eta_{XY} + C_b^{(2)} {(g^*)}^{2} + \cdots,\\
\eta_f &=& \eta_{\psi} = C_f^{(2)} {(g^*)}^{2} + \cdots,   
\end{eqnarray}
where we have calculated $C_f^{(2)} \simeq -0.03$. The anomalous dimension
for the 3d XY model (appropriate for one vortex species) has been
calculated by Hasenbusch and T\"{o}r\"{o}k using Monte Carlo methods
\cite{HT}; they find $\eta_{XY} \simeq 0.038$. 
The anomalous dimension will also enter into the spin-spin correlation
function. Within our model, it looks as though vertex correction
diagrams will not contribute as much near the critical point as the
direct ${[G_{\psi}]}^{2}$ term.
 
\section{conclusions}

In this paper, we have used a gauge theory of strongly-interacting
electrons to explore the regions near the superconducting state in the
high-$T_c$ cuprates. This gauge theory 
exhibits spin-charge separated and spin-charge confined phases. 
We have seen that the presence of one electron per unit cell has
profound implications for the regions near the superconducting state.
Within a dual description, half-filling of electrons corresponds to
fully-frustrated vortices, leading to a spontaneous breaking of
translational symmetries in the electron system. From this, we
have seen the possibility of striped superconductivity as well as a
host of confined insulators descending from d-wave
superconducting phases. We have then used Chern-Simons
methods to calculate lowest-order critical properties of the
confinement transition between these phases. Because we have worked at
half-filling of electrons throughout, our results are of
particular relevance to the undoped cuprate materials, which may be
spin-charge confined. However, we also hope that the flavor of our
results may be of interest in the heavily overdoped materials, where
the confinement of spin and charge may result in a Fermi liquid phase.

We are grateful to Leon Balents, Patrick Lee, and Doug
Scalapino for insightful discussions. This research was generously
supported by the NSF under Grants DMR-97-04005, DMR95-28578 and
PHY94-07194.

\end{multicols}

\begin{references}
\bibitem{expU1} In our action, the presence of spin-charge separation
is dictated by the phase of the $Z_2$ gauge field, which, unlike the
SU(2) and U(1) gauge theories considered by other authors, exhibits both
confining and deconfining phases.
\end{references}
\end{document}